\documentstyle[graphicx,prl,aps]{revtex} 
   \newcommand{\cent}[1] {\begin{center}#1\end{center}}
   \newcommand{\doublint}{\int\rule{-3.5mm}{0mm}\int}
   \newcommand{\mra}  {\to}

   \newcommand{\vecb}[1]{\mbox{\bf#1}}
\begin{document}
\paperwidth =15cm
\draft
\title{Phase Transitions in ``Small'' Systems.}
\author{D.H.E. Gross and E.Votyakov} \address{ Hahn-Meitner-Institut
  Berlin, Bereich Theoretische Physik,Glienickerstr.100\\ 14109
  Berlin, Germany and Freie Universit{\"a}t Berlin, Fachbereich
  Physik; \today} \maketitle
\begin{abstract} 
  Traditionally, phase transitions are defined in the thermodynamic
  limit only. We discuss how phase transitions of first order (with
  phase separation and surface tension), continuous transitions and
  (multi)-critical points can be seen and classified for {\em small}
  systems. ``Small'' systems are systems where the linear dimension is
  of the characteristic range of the interaction between the
  particles.  I.e.  also astrophysical systems are ``small'' in this
  sense.  Boltzmann defines the entropy as the logarithm of the area
  $W(E,N)=e^{S(E,N)}$ of the surface in the mechanical N-body phase
  space at total energy $E$.  The topology of $S(E,N)$ or more
  precisely, of the curvature determinant $D(E,N)=\partial^2S/\partial
  E^2*\partial^2S/\partial N^2-(\partial^2S/\partial E\partial N)^2$
  allows the classification of phase transitions {\em without taking
    the thermodynamic limit}.  Micro-canonical thermo-statistics and
  phase transitions will be discussed here for a system coupled by
  short range forces in another situation where entropy is not
  extensive.  The first calculation of the {\em entire entropy surface
    $S(E,N)$} for the diluted Potts model (ordinary ($q=3$)-Potts
  model plus vacancies) on a $50*50$ square lattice is shown. The
  regions in \{$E,N$\} where $D>0$ correspond to pure phases, ordered
  resp.  disordered, and $D<0$ represent transitions of first order
  with phase separation and ``surface tension''. These regions are
  bordered by a line with $D=0$.  A line of continuous transitions
  starts at the critical point of the ordinary ($q=3$)-Potts model and
  runs down to a branching point $P_m$. Along this line
  $\mbox{\boldmath $\nabla$}D$ vanishes in the direction of the
  eigenvector $\mbox{\boldmath $v_1$}$ of $D$ with the largest
  eigen-value $\lambda_1\approx 0$. It characterizes a maximum of the
  largest eigenvalue $\lambda_1$.  This corresponds to a critical line
  where the transition is continuous and the surface tension
  disappears.  Here the neighboring phases are indistinguishable. The
  region where two or more lines with $D=0$ cross is the region of the
  (multi)-critical point.  The micro-canonical ensemble allows to put
  these phenomena entirely on the level of mechanics.
\end{abstract}
\section{Introduction}
Nuclei, atomic clusters and astrophysical objects are not large
compared to the range of their forces.  Therefore, these systems are
not extensive.  (In the following we call systems ``non-extensive'' in
a somewhat more general sense: If they are divided into pieces, their
entropy is not the sum of the entropies of their parts in contrast to
conventional extensive systems where this is assumed at least if the
pieces are themselves macroscopic).  Although the largest systems
possible belong to this group we call these systems ``small''. But
also at phase transitions in systems with short range forces does the
entropy of the surface separating the different phases not scale with
the size of the system. Lateron we {\em define systems to be
  ``small'' precisely by the condition that the entropy $S(E,N,V)$
  does not scale with the number of particles or the volume and
  especially if $S(E,N,V)$ has some convex regions.}

Most applications of thermodynamics to ``small'' systems are more or
less transcriptions from the thermodynamics of macroscopic systems,
c.f.  the book by Hill \cite{hill64}. Conventional thermo-statistics,
however, relies heavily on the use of the thermodynamic limit
($V\!\!\mra\!\!\infty|_{\mbox{\scriptsize{$N/V$, or $\mu$ const.}}}$)
and extensivity, c.f. e.g. the book of Pathria \cite{pathria72}. This
is certainly not allowed for our systems.  Extensivity is nowadays
considered to be an essential condition for thermodynamics to work,
c.f.  Lieb and Ygnvason \cite{lieb97}.  That the {\em micro-canonical}
statistics works well also for ``small'' systems without invoking
extensivity will be demonstrated here for finite normal systems. The
use of the thermodynamic limit and of extensivity, however, is closely
intervowen with the development of thermodynamics and statistical
mechanics since its beginning more than hundred years ago.  When we
extend thermodynamics to ``small'' systems we should establish the
formalism of thermodynamics starting from mechanics in order to remain
on a firm basis. This is an old program since Boltzmann and Einstein
\cite{boltzmann1877,einstein04}.  We will see how this idea guides us
to more and deeper insight into the most dramatic phenomena of
thermodynamics, phase transitions. Moreover, it gives the most natural
extension of thermo-statistics to some non-extensive systems without
invoking any modification of the entropy like that proposed by Tsallis
\cite{tsallis88} see also \cite{raggio99}.

In the following section we sketch a deduction of thermo-statistics
from the principles of mechanics alone. Nothing outside of mechanics
must be invoked. This was the starting point of Boltzmann
\cite{boltzmann1877} , Gibbs \cite{gibbs02}, Einstein
\cite{einstein03,einstein04} and the Ehrenfests
\cite{ehrenfest12,ehrenfest12a} at the beginning of this century.
They all agreed on the logical hierarchy of the micro-canonical as the
most fundamental ensemble from which the canonical, and
grand-canonical ensembles can be deduced under certain conditions.
According to Gibbs the latter two approximate the micro ensemble in
the thermodynamic limit of infinitely many particles if the system is
homogeneous. Then surface effects and fluctuations can be ignored
relatively to the bulk mean values. This is the main reason why the
thermodynamic limit became basic in the statistical foundation of
macroscopic thermodynamics.  However, it was Gibbs \cite{gibbs02f} who
stressed that the equivalence of the three ensembles is not even true
at phase transitions of first order, even in the thermodynamic limit.

This chapter repeats the familiar deduction of statistical mechanics
as it was first formulated by Gibbs \cite{gibbs02}.  However, we will
indicate at which point the main stream of logical steps towards a
thermodynamics of ``large'' systems has to be left in order to cover
also ``small'' and non-extensive systems.

Just a remark is neccessary here: One might think that a small system
embedded in a heat bath may be described by the canonical or even the
grand-canonical ensemble. A heat bath interacts with the system via
its surface. For small systems these surface mechanisms are important.
They affect the physical properties and must be specified.  Certainly
more must be controlled in such cases than just the mean energy
(temperature) as is assumed in a canonical treatment. This would be
only correct for an infinite system, with short range interactions,
and if we wait an infinitely long time, then the details of the
physical mechanisms in the surface contacting the system with the bath
become unimportant relatively to the bulk.

The third section addresses phase transitions. First, the basics of the
conventional definition of phase transitions in the thermodynamic
limit ($V\!\!\mra\!\!\infty|_{\mbox{\scriptsize N/V, or $\mu$
    const.}}$) by the theory of Yang and Lee \cite{lee52} are
reminded. Then this definition is translated into the micro-canonical
picture and the source of the non-analyticities of the canonical
thermodynamic potentials is traced to the topology of the
micro-canonical entropy function, esp. its curvature.

In the fourth section we explain the general features of the
micro-canonical phase diagram as function of fixed energy density
($e=E/V$) and particle number- ($n=N/V$)/magnetization-density, at first,
with the well known example of the Ising model and the ($q=1$) diluted
Potts model.  It is shown how its phase space is limited by the line
of random configurations which have the maximum possible entropy.
Further the region of ``field driven'' phase separation into
spontaneously magnetized clusters (first order transition) can be well
studied in the energy-magnetization or the energy-particle number
phase diagram.  (Remark: the distinction between ``field driven'' and
``temperature driven'' transition makes sense only for the Ising
model. Due to its intrinsic symmetry $M\to -M$ the energy- axis is
also the main curvature direction for $M=0$. This is of course not the
case for systems without this symmetry.)  Then the diluted Potts model
with $q=3$ is introduced.  It is sketched which features of its phase
diagram are to be expected in the case of a ``small'' system.

Section V gives a short discussion of the main steps of the simulation
and in the section VI the numerical results are shown. We present the
first view of the entire entropy-density surface $s(e,n)$ as function
of energy density ($e$) and particle number density ($n$) for the
diluted ($q=3$)-Potts model on a {\em finite} 2-dim lattice with
``volume'' $V=L^2$.\footnote{In the following we often skip the volume
  $V$ as third variable as this is always hold fixed. Of course for
  finite systems the {\em shape} of the container matters. In the
  diluted Potts model we use a square lattice with periodic boundaries
  (``square torus'').} More insight into the phase diagram of the
model is given by an overall view of the determinant of the curvature
density $d(e,n)$.  Various kinds of phase transitions and critical
lines can be seen and classified. A subsection is addressed to the
essential differences between the micro-canonical and the
grand-canonical phase diagram.  The information lost by the latter
representation is emphasized.

The following section discusses the validity of the second law of
thermodynamics in non-extensive systems with a convex $s(e,n)$. The
consequences of convexity of the entropy for Weinhold's geometric
interpretation of thermodynamics are discussed. It will be explained
that a convexity of $S(E,N)$ is not in contradiction to the second law
of thermodynamics.

Finally, our main conclusions on phase transitions in ``small''
non-extensive systems are summarized in section VIII.
\section{On the mechanical background of thermo-statistics}

An equilibrated many-body system is characterized by few macroscopic
quantities:
\begin{enumerate}
\item Its energy $E$, mass (number of atoms) $N$, volume $V$,
\item its entropy $S$,
\item its temperature $T$, pressure $P$, and chemical potential $\mu$.
\end{enumerate}
There are important qualitative differences between these three
groups: All variables of the first group have a clear mechanical
significance. They are conserved and well defined at each point of the
N-body phase space. The internal dynamics of the system cannot leave
the shell in phase space which is defined by these variables.  Also
entropy as the most important quantity within thermodynamics has a
clear mechanical foundation since Boltzmann.  His gravestone has the
famous epitaph: \cent{\fbox{\fbox{S=k$\cdot$ lnW}}} relating the entropy $S$
to the size $W(E,N,V)=\epsilon_0 tr\delta(E-H_N)$ of the energy ($E$)
surface in the N-body phase space at given volume ($V$).  Here
$\epsilon_0$ is a suitable small energy constant which does not affect
any variation of the entropy, $H_N$ is the $N$-particle Hamiltonian,
and
\begin{equation}
tr\delta(E-H_N)=\int{\frac{d^{3N}p\;d^{3N}q}{(2\pi\hbar)^{3N}}
\delta(E-H_N\{q,p\})}.
\end{equation}
The set of points on this surface defines the micro-canonical
ensemble.  In contrast to the conserved quantities which are defined
at each phase space point, the entropy refers to the whole
micro-canonical ensemble.

Remark: For a system with discrete energies $E_i$ e.g. a lattice or a
quantum system with energy spacing $\epsilon_0$ one should define the
micro-canonical partition sum by the number of states at this energy.
When we discuss derivatives of $W$ we imagine a suitable smoothing of
this.

It is important to notice that Boltzmann's and also Einstein's
formulation allows for {\em defining the entropy entirely within
  mechanics} by $S_{micro}:=ln[W(E,N,V)]$. It is a single valued,
non-singular, in the classical case multiply differentiable, function
of all {\em ``extensive'', conserved} dynamical variables. No
thermodynamic limit must be invoked and this definition applies to
non-extensive like our ``small'' systems as well.
 
The third group of quantities which characterize the thermodynamical
state of an equilibrated many-body system, temperature, pressure and
chemical potential have no immediate mechanical significance. Within
micro-canonical statistics they are {\em defined} by the derivatives
of the entropy $S(E,N,V)$ vs. the conserved quantities :
\begin{eqnarray}
\frac{1}{T}&=&\frac{\partial S}{\partial E}\\
\mu&=&-T\frac{\partial S}{\partial N}\\
P&=&T\frac{\partial S}{\partial V}.
\end{eqnarray} From the mechanical point of view they are secondary,
derived quantities. This difference to the two other groups of
variables will turn out to be significant for ``small'' systems.
Again, like entropy itself, these quantities characterize the whole
micro-canonical ensemble, not an individual point in the N-body phase
space.

Starting from this point, the conventional thermo-statistics assumes
extensivity and explores the thermodynamic limit
($V\!\!\mra\!\!\infty|_{\mbox{\scriptsize N/V, or $\mu$ const.}}$)
c.f.  \cite{pathria72}. This procedure follows Gibbs \cite{gibbs02}.
He introduced the canonical ensemble, which since then is the basic of
all modern thermo-statistics. The link between both ensembles is
established by a Laplace transform. E.g. the usual grand-canonical
partition sum is the double Laplace transform of the micro-canonical
partition sum $W(E,N,V)=e^{S(E,N,V)}$:
\begin{equation}
Z(T,\mu,V)=\doublint_0^{\infty}{\frac{dE}{\epsilon_0}\;dN\;e^{-[E-\mu
N-TS(E,N,V)]/T}}=\frac{V^2}{\epsilon_0}\doublint_0^{\infty}{de\;dn\;e^{-V[e-\mu
n-Ts(e,n,V)]/T}}\label{grandsum}.
\end{equation}
In the thermodynamic limit it is useful to work with the energy
density $e=E/V$, the particle number density $n=N/V$, and the entropy
density $s=S/V $, cf. the right expression in eq.(\ref{grandsum}).
\section{Phase transitions micro-canonically}

According to Yang and Lee \cite{lee52} phase transitions are indicated
by singularities of the grand-canonical potentials
($\propto\frac{1}{V}ln[Z]$) as function of $z=e^{\mu/T}$ on the
positive real $z$-axis.  These, however, can occur in the
thermodynamic limit ($V\!\!\mra\!\!\infty|_{\mbox{\scriptsize $\mu$
    const.}}$) only. For finite volumes the number of particles $N$ is
finite.  Consequently, $Z$ is a sum of a finite number of powers $z^N$
and $\frac{1}{V}ln[Z]$ is analytical for positive $z$ at any $T$. Are
there no phase transitions in finite systems?

There are phenomena observed in finite systems which are typical for
phase transitions. Sometimes this is even so in astonishingly small
systems like nuclei and atomic clusters of $\sim 100$ atoms
\cite{gross95,schmidt97,haberland99}. In ref.\cite{gross157} we showed
that their characteristic parameters as transition temperature, latent
heat, and {\em surface tension} are -- in the case of some metals --
already for thousand atoms close, though of course not equal, to their
known bulk values.  Therefore, it seems to be fully justified to speak
in these cases of phase transitions of first order.

We need an extension of thermodynamics to ``small'' systems which
avoids the thermodynamic limit.  However, here is a severe problem.
{\em The three popular ensembles, the micro-canonical, the canonical,
  and the grand-canonical ensembles are not equivalent for ``small''
  systems.} The energy {\em per particle} can fluctuate around its
mean value $<\!\!E/N\!\!>$ in the (grand-)canonical ensemble whereas
the energy fluctuations are zero in the micro-canonical ensemble.
Moreover, the heat capacity is strictly positive in the canonical
ensembles whereas it may become {\em negative} in the micro ensemble.
It was Gibbs himself who warned for the use of the canonical ensemble
at phase transitions of first order \cite{gibbs02f}. In this situation
it is certainly advisable to keep close contact with mechanics.  It is
helpful to realize that the fundamental micro-canonical ensemble as
introduced by Boltzmann is the only one which has a clear mechanical
definition \cite{boltzmann1884,einstein04} for finite systems.

To extend the definition of phase transitions by Yang and Lee to
finite systems we study which feature of the micro-canonical partition
sum $W(E,N,V)$ leads to singularities of the grand-canonical
potentials $\frac{1}{V}ln[Z]$ as function of $z=e^{\mu/T}$ by the
Laplace transform eq.(\ref{grandsum}). In the thermodynamic limit
$V\to\infty|_{\mbox{\scriptsize $\mu$ const.}}$ this integral can be
evaluated by asymptotic methods. As far as the entropy surface
$s(e,n)$ has everywhere negative curvatures the integrand of eq.
(\ref{grandsum}) has a single maximum.  For large $V$ the Laplace
integral (r.h.s of eq.(\ref{grandsum})) is then dominated by the
contribution of this peak. This is a {\em stationary} point
\{$e_s$,$n_s$\} where $T^{-1}=\partial s/\partial e$,
$\nu=-\mu/T=\partial s/\partial n$. If there is only a single
stationary point then there is a one to one mapping of the
grand-canonical ensemble to the micro-canonical one and
energy-fluctuations disappear.

This, however, is not the case at phase transitions of first order.
Here the grand-canonical ensemble contains several Gibbs states
(stationary points) at the same temperature and chemical potential
which contribute similarly to the integral (\ref{grandsum}).
Consequently, the statistical fluctuations of $e$ and $n$ do not
disappear in the grand-canonical ensemble even in the thermodynamic
limit. This is the reason why Gibbs himself excluded phase separations
\cite{gibbs02f}.  Between the stationary points {\em $s(e,n)$ has at
  least one principal curvature $\geq 0$}.  Here van Hove's concavity
condition \cite{vanhove49} for the entropy $s(e,n)$ is violated. In
the thermodynamic limit these points get jumped over by the integral
(\ref{grandsum}) and $ln[Z]$ becomes non-analytic.  Consequently, we
{\em define phase transitions also for finite systems topologically by
  the points and regions of non-negative curvature of the entropy
  surface $s(e,n)$ as a function of the mechanical, conserved
  ``extensive'' quantities like energy, mass, angular momentum etc.}.

The central quantity of our further discussion the determinant of the
curvatures of $s(e,n)$ is defined as
\begin{equation}
d(e,n)= \left\|\begin{array}{cc}
\frac{\partial^2 s}{\partial e^2}& \frac{\partial^2 s}{\partial n\partial e}\\
\frac{\partial^2 s}{\partial e\partial n}& \frac{\partial^2 s}{\partial n^2}
\end{array}\right\|
= \left\|\begin{array}{cc}
s_{ee}&s_{en}\\
s_{ne}&s_{nn}
\end{array}\right\|. \end{equation}

Also critical fluctuations, i.e. abnormally large fluctuations of some
extensive variable in the grand-canonical ensemble or the eventual
divergence of some susceptibilities are micro-canonically connected to
the vanishing of the curvature determinant, e.g. in the following
examples of $d(e,n)$ or $d(e,m)$ respectively:
\begin{eqnarray}
&&\mbox{The micro-canonical specific heat is given by :}\nonumber\\
c_{micro}(e,n,V)&=&\left.\frac{\partial e}{\partial T}\right|_\nu
=-\frac{s_{nn}}{T^2d(e,n)}\;,\\
d&=&\frac{d (\beta\nu)}{d(en)}\\
&&\mbox{or the isothermal magnetic susceptibility by :}\nonumber\\
\chi_{micro,T}(e,n,V)&=&\left.\frac{\partial m}{\partial B}\right|_T
=\frac{s_{ee}}{d(e,m)}\;,\\
&&\mbox{with }s_{ee}=\frac{\partial^2 s}{\partial e\partial e}\mbox{ etc.}
\end{eqnarray}
In the case of a classical continuous system $s(e,n)$ is everywhere
finite and multiply differentiable. In that case the inverse
susceptibilities $[c_{micro}(e,n,V)]^{-1}$ and $[\chi_{micro,
  T}(e,n,V)]^{-1}$ are well behaved smooth functions of their
arguments even at phase transitions.  Problems arise only if the
susceptibilities are considered as functions of the ``intensive''
variables $T$, and $\nu$ or $B$ \cite{straley73}.  In the case of
lattice systems we can only assume that the inverse susceptibilities
are similarly well behaved.  This will be further illuminated in the
section \ref{lostinfo}.

Experimentally one identifies phase transitions of first order of
course not by the non-analyticities of $\frac{1}{V}ln[Z]$ but by the
interfaces separating coexisting phases, e.g. liquid and gas, i.e. by
the {\em inhomogeneities} of the system which become suppressed in the
thermodynamic limit in the grand-canonical ensemble. This fact was
early realized by Gibbs \cite{gibbs06} and he emphasized that using
$S$ vs.  volume at phase separation ``has a substantial advantage over
any other method because it shows the region of simultaneous
coexistence of the vapor, liquid, and solid phases of a substance, a
region which reduces to a point in the more usual pressure-temperature
plane.''  That is also the reason why for the grand-canonical ensemble
the more mathematical definition of phase transitions \cite{lee52} is
needed.  The main advantage of the micro-canonical ensemble is that it
allows for {\em inhomogeneities} as well and we can keep much closer
to the experimental criteria.

Interfaces have three opposing effects on the entropy:
\begin{itemize}
\item An entropic {\em gain} by putting a part ($N_1$) of the system
  from the majority phase (e.g. solid) into the minority phase
  (bubbles, e.g. gas) with a higher entropy per particle. However,
  this has to be paid by additional energy $\Delta E$ to break the
  bonds in the ``gas''-phase. As both effects are proportional to the
  number of particles $N_1$ being converted, this part of the entropy
  rises linearly with the additional energy.
\item With rising size of the bubbles their surfaces grow. This is
  connected to an entropic {\em loss} (surface entropy) proportional
  to the interface area due to additional correlations between the
  particles at the interface. As the number of surface atoms is
  $\propto N_1^{2/3}$ this is not linear in $\Delta E$ and leads to a
  convex intruder in $S(E,N,V)$, the origin of surface tension
  \cite{gross150}.\label{surf}
\item An additional mixing entropy for distributing the
  $N_1$-particles in various ways over the bubbles.
\end{itemize}
At a (multi-) critical point two (or more) phases become
indistinguishable because the interface entropy (surface tension)
disappears.
\section{The model and its qualitative features}
In this paper we want to discuss how critical and also multi-critical
points manifest themselves in the micro-canonical statistics of a
finite diluted Potts model.

It is helpful first to remind the main physical effects to be
expected: We start with the well known properties of the Ising model
(written here as a Potts model with $q=2$ \cite{wu82}). The
Hamiltonian and the magnetization are defined as:
\begin{eqnarray}
H&=&-\sum_{i,j}^{n.n.pairs}
{\delta_{\sigma_i,\sigma_j}} -B\sum_i{\sigma_i}\\
M&=&\sum_i{\sigma_i}.
\end{eqnarray}The double sum runs over nearest neighbor spin-pairs only
\cite{wu82,griffith72}.

The two-dimensional Ising model was extensively discussed as function
of temperature $T$ and magnetic field $B=0$ in the grand-canonical
ensemble for $L\to \infty$ (thermodynamic limit)
\cite{griffith72,pathria72}.  As known from Onsager's solution it has
a continuous (second order) critical point at $B=0$ and
$\beta_c=\frac{1}{T_c}=0.5\ln(\sqrt{2}+1)$\cite{wu82,lavis99}. Above
the critical point, $T>T_c$ there is a disordered phase with vanishing
magnetization for $B=0$.  Below $T_c$ there is a bifurcation of the
spontaneous magnetization curve $M(T,B\to
0)=\mbox{sign}(B)L^2m_{spontan}(T)$. As function of the extensive
variables \{$E,M$\} between $M=\pm L^2m_{spontan}(T)$ there is a whole
region with phase separation which is inaccessible to the
grand-canonical ensemble. Also the internal energy per lattice point
$e(T,B=0)$ is analytically known in the thermodynamic limit c.f.
\cite{pathria72}.  This may be inverted to get the function:
\begin{equation}
m(e,B\to 0)=\left\{\begin{array}{cl}
0&,e\ge e_c=-(1+\frac{1}{\sqrt{2}})\\
\mbox{sign}(B) m_{spontan}(e)&, e\le e_c\\
\end{array}\right. .
\end{equation}

Then the phase diagram in the \{$e,m$\} plane can be drawn as in
fig.\ref{ising}:
\begin{center}\begin{figure}
    \includegraphics*[bb = 0 0 290 180, angle=-0,
    width=11cm,clip=true]{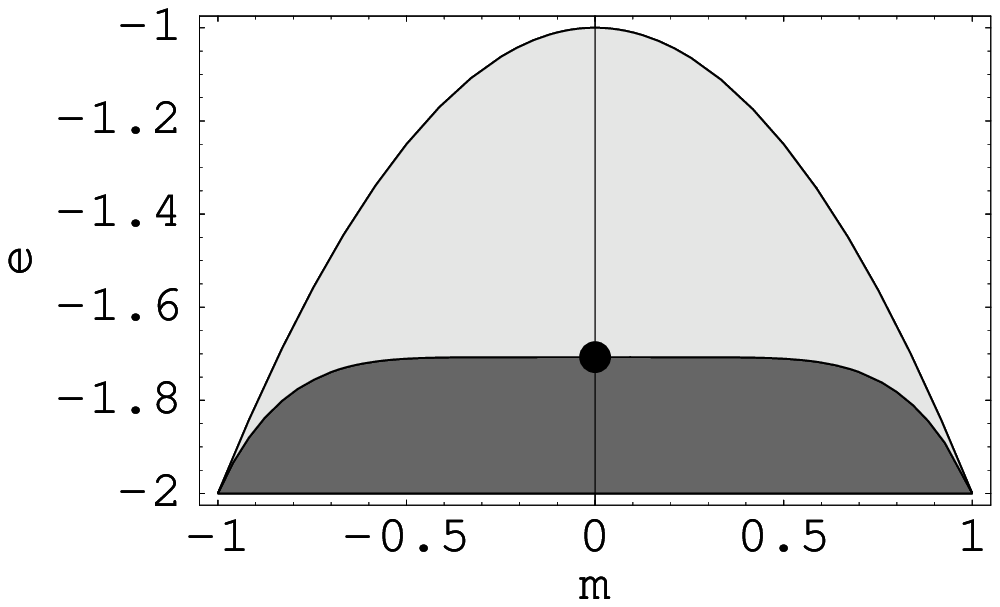}
    \\
\caption{Phase diagram in magnetization ($m$) and energy ($e$) per 
  spin particle for the Ising model (ordinary ($q=2$)-Potts model) in
  the thermodynamic limit. The upper parabolic line is the locus of
  the completely random configuration $e=-(m^2+1)$ with maximum
  entropy. The physical possible phase space points are in the two
  shaded regions.  The darkly shaded region is the phase space of
  first order transition and coexistence of drops with positive and
  drops with negative magnetization. It is inaccessible in the
  grand-canonical ensemble. The dot gives the critical point
  ($e_c=-(1+\frac{1}{\sqrt{2}})$, $m_c=0$).
\label{ising}}
\end{figure}\end{center}

At energies $e>e_c=-(1+\frac{1}{\sqrt{2}})$ one has a disordered
(``gas'') phase. At $e_c=-(1+\frac{1}{\sqrt{2}}),m=0$ there is a
critical point of continuous transition with vanishing surface
tension. From here an {\em inhomogeneous} region of condensed phase
starts towards lower energies (several clusters of positive or
negative magnetization coexist separated by interfaces). In the
thermodynamic limit the overall magnetic field $<B>$ vanishes here.
This region is bordered by the curve $m_{spontan}(e,B=0)$.
Approaching this border from inside with fixed $e$ and rising $m$, the
clusters with the opposite magnetization get depleted and the surfaces
between the clusters disappears. (However, even when there are no
clusters of the wrong magnetization there should still be {\em
  fluctuations} with the other, wrong magnetization because these
states have a nonvanishing entropy. There is a general impossibility
to distinguish a phase separation from a fluctuation within an
individual configuration. This is a realistic example for the
definition of a phase as a property of the entire ensemble only.)
This region of phase separation is darkly shaded in fig.\ref{ising}
and is inaccessible to the grand-canonical ensemble.

The following Gedanken-experiment may illuminate why this region
cannot be accessed in the grand-canonical ensemble: Suppose one
prepares the system at a sharp energy and magnetization in the region
with positive curvature. Then one puts the system into a heat bath
with temperature $T_{tr}$ and magnetization (particle-number in the
case of the diluted Potts model) bath with $\mu_{tr}$ where it will
be suddenly exposed to energy- and magnetization- (particle number)
fluctuations of arbitrary size. Because of the positive curvature of
its entropy the system is unstable and will receive or loose
sufficient energy to convert {\em entirely} into the gas or {\em
entirely} into the solid phase where its free energg is lower. The
latent heat which is neccessary for this will be spended (absorbed)
by the bath.

\begin{center}\begin{figure}
    \includegraphics*[bb = 0 0 290 180, angle=-0,
    width=11cm,clip=true]{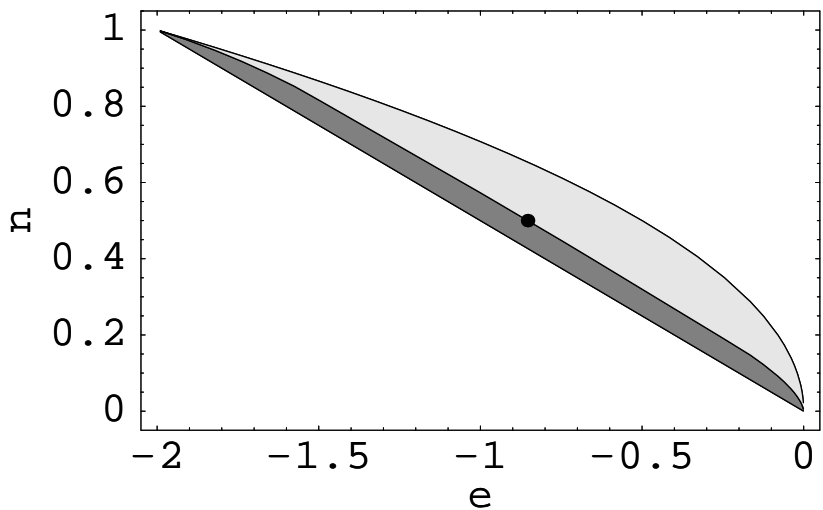}
    \\
\caption{Phase diagram for the diluted ($q=1$)-Potts model in the
  thermodynamic limit. With the transformation $\sigma_i=2o_i-1$ of
  the two possible occupation numbers to the spin on site $i$ and
  $m=\frac{1}{L^2}\sum_i{\sigma_i}$ this is analogous to the
  energy-magnetization plot of the Ising model. The dot gives the
  critical point ($e_c\approx-0.853$, $n_c=0.5$). The darkly shaded region
  is the phase space of phase coexistence which is inaccessible in the
  grand-canonical ensemble. \label{isingG}}
\end{figure}\end{center}

The Ising model can be modified by the transformation
$o_i=\frac{1}{2}(\sigma_i+1)$ and
$H_{lg}=-\frac{1}{2}\sum_{i,j}{\delta_{o_i,+1}\delta_{o_j,+1}}$ into
the diluted ($q=1$)-Potts model. Here the lattice points with $o_i=0$
are vacancies. Figure (\ref{isingG}) shows the phase diagram of the
diluted ($q=1$)-Potts model corresponding to the phase diagram of the
Ising model as discussed above. The three lines indicate (from left to
right) the ground state ($e_0(n)=-2n$), the critical line, and the
line of random (maximum entropy) configuration ($e_{max}(n)=-2n^2$).
The dot gives the critical point which corresponds to the critical end
point of the phase separation in the liquid--gas transition.  Between
the ground state and the critical line is the \{$e,n$\}-region (darkly
shaded) of first order phase transition with phase separation. Here
the system prefers {\em inhomogeneous} configurations with droplets
enbedded in the gas. This corresponds to the above mentioned similar
inhomogeneous region of the Ising model at energies below the Ising
critical point ($e_c=-(1+\frac{1}{\sqrt{2}})$,$m_c=0$), where clusters
of positive magnetization and negative magnetization coexist. In the
thermodynamic limit we expect the micro-canonical entropy $s(e,n)$ to
be flat in the shaded region. The whole darkly shaded region of
inhomogeneities is not accessible in the grand-canonical ensemble.
This is a striking example of the dramatic loss of information when
the grand-canonical ensemble is addressed to phase transitions.  Above
the critical line we have the region of the pure disordered phase.
Here $s(e,n)$ is concave.

In the following we investigate the 3-states diluted Potts model now
on a {\em finite} 2-dim (here $L^2=50^2$) lattice with periodic
boundaries in order to minimize effects of the external surfaces of
the system. The model is defined by the Hamiltonian:
\begin{eqnarray}
H&=&-\sum_{i,j}^{n.n.pairs}o_i o_j\delta_{\sigma_i,\sigma_j}\\
n&=&L^{-2}N=L^{-2}\sum_io_i .\nonumber
\end{eqnarray}
Each lattice site $i$ is either occupied by a particle with spin
$\sigma_i =1,2,\mbox{ or }3$ or empty (vacancy).  The sum is over
neighboring lattice sites $i,j$, and the occupation numbers are:
\begin{equation}
o_i=\left\{\begin{array}{cl}
1&\mbox{, spin particle in site }i\\
0&\mbox{, vacancy in site }i\\
\end{array}\right. .
\end{equation}

This model is an extension of the ordinary ($q=3$)-Potts model to
allow also for vacancies. At zero concentration of vacancies ($n=1$),
the system has a continuous phase transition at
$e_c=1+\frac{1}{\sqrt{q}}\approx 1.58$\cite{baxter73,pathria72}. With rising
number of vacancies the probability to find a pair of particles at
neighboring sites with the same spin orientation decreases.  The
inclusion of vacancies has the effect of an increasing effective
$q_{eff}\ge 3$. This results in an increase of the critical energy of
the continuous phase transition with decreasing $n$ and provides a
line of continuous transition, which is supposed to terminate when
$q_{eff}$ becomes larger than $4$. Here the transition becomes first
order.

At smaller energies the system is in one of three ordered phases
(spins predominantly parallel in one of the three possible
directions).  We call this the ``solid'' phase. This scenario gets
full support by our numerical findings in the next sections.

In the following, we will show for the case of the diluted
($q=3$)-Potts model how the total micro-canonical entropy surface
$S(E,N)$ uncovers first order phase transitions, continuous phase
transitions, critical, and multi critical points even for small
systems and non-extensive systems.  This is an extension of previous
work on systems with one thermodynamic degree of freedom like the
fragmentation phase transition in highly excited nuclei
\cite{gross95}, in atomic clusters \cite{gross157}, and in the
ordinary Potts model \cite{gross150,hueller94}.  The main purpose for
the study of systems with two thermodynamic degrees of freedom like
\{$E,N$\} is the possibility to localize (multi-)critical points.
\section{Simulation method}

The simulation methods proceeded in two steps: We first covered all
space $\{E=e*L^2,N=n*L^2\}$, $L=50$ by a mesh with about $1000$ knots
with distances of $\Delta e=0.08$ and $\Delta n=0.04$.  At each knot
$\{e_i,n_k\}$ we performed micro-canonical simulations ($\approx
2*10^8$ events) to get a histogram of the probabilities
$P(e_i,n_k)=e^{S}$ for the system to be distributed in the narrow
region $(E_i\pm 4)*(N_k\pm 4)$ of phase space.  Local derivatives
$\beta=\left(\partial S(E,N)/\partial E\right)_N$,
$-\beta\mu=\nu=\left(\partial S(E,N)/\partial N\right)_E$ in each
histogram give a first ``measurement'' of the ``intensive''
quantities.  These are used to interpolate $\beta,\nu$ over the gaps
between the knots of our mesh c.f. \cite{gross170}.
\begin{enumerate}
\item From the first interpolation we get an estimate $\beta_0,\nu_0$
  for the center of any region \{$\Delta e*\Delta n$\}.
\item In a second iteration step we sample
  $P_1(e,n)=P(e,n)*e^{-\beta_0 L^2(e-\mu_0 n)}$ in a broader \{$e,n$\}
  domain.  This is chosen to touch or overlap the next neighboring
  region. This way a dense covering of the entire \{$e,n$\} space is
  achieved.
\item A fourth order regression fit $f_4(e,n)$ in $e$ and in $n$ to
  $\ln[P_1(e,n)]$ is performed, from which we determine the two
  principal axes of curvature $\lambda_1, \lambda_2$ of
  $s(e,n)$ and then can rotate if desired the frame of the basic
  rectangle in \{$e,n$\} to be parallel to the main curvature axes.
\item The last step is then a longer and extensive sampling of
  $P_2(e,n)= e^{-f_4(e,n)}P_1(e,n)$ in the new basic rectangle.
\item Eventually one has to recycle steps $3$ and $4$ a few times up
  to the time one obtains a sufficiently uniform covering of
  $P_2(e,n)$.
\item At the end on gets back the original by $P(e,n)=e^{f_4(e,n)}
  e^{\beta_0 L^2(e-\mu_0 n)}P_2(e,n)$
\end{enumerate} 
We obtain a broad and uniform covering of the whole \{$e,n$\} space.
This way one gets the most effective (and fast) ``measurement'' of
$s(e,n)$, $\beta(e,n)$, $\nu(e,n)$ up to all third order derivatives
of $s(e,n)$ e.g. $s_{ee}(e,n)$ or $s_{een}(e,n)$ etc..  Technical
details of our method will be published in \cite{gross169}.

\section{Results}
\begin{center}\begin{figure}
    \includegraphics*[bb = 0 0 295 265, angle=-0,
width=15cm,clip=true]{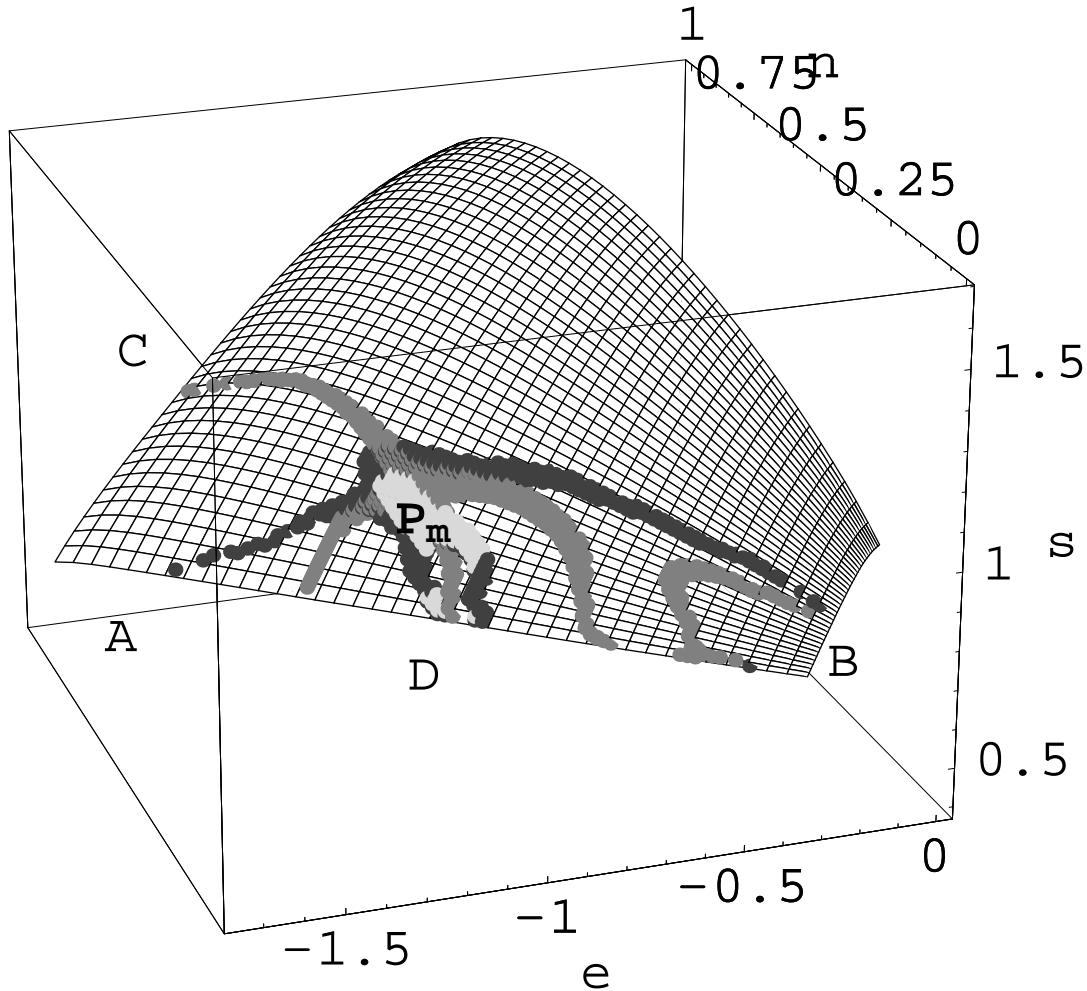}
\\
\caption{ Entropy $s_{micro}(e,n)$.\label{6d3_e}}
\end{figure}\end{center}

\begin{figure}
  \includegraphics*[bb =0 0 290 180, angle=90, width=15cm,
clip=true]{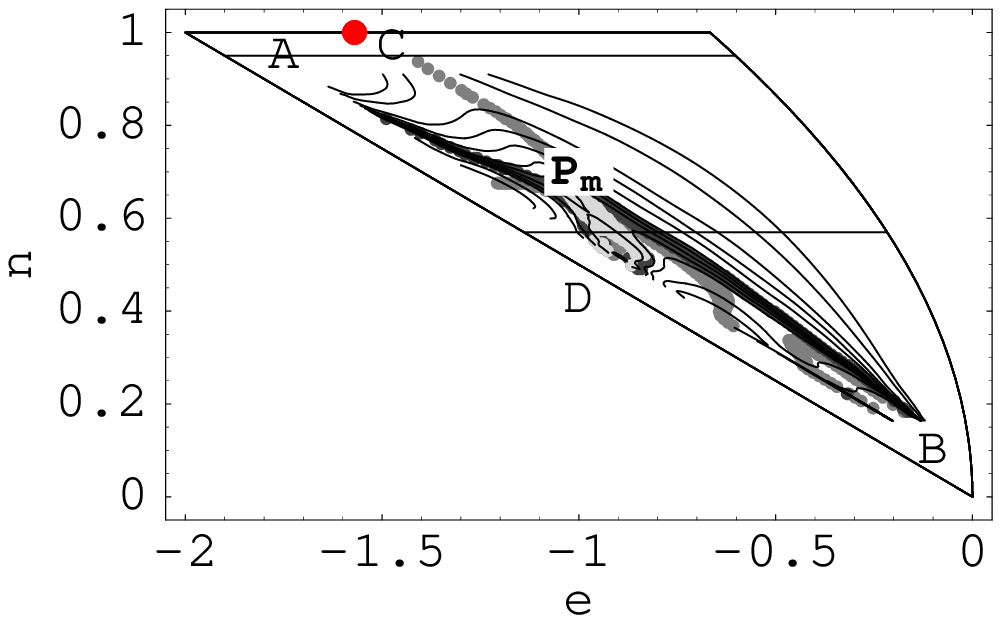}
\\
\caption{Contour plot of the determinant of curvatures
  $d(e,n)$ defined in eq.(\ref{curvdet}).The grey levels are as in
  fig.\ref{6d3_e}: Regions above $\widehat{CP_mB}$ : concave, $d>0$,
  pure phase (disordered, gas), in the triangle $A$$P_m$$C$ concave,
  pure phase (ordered, solid); below $\widehat{AP_mB}$: convex, $d<0$,
  phase-separation, first order; At the dark lines $\widehat{AP_mB}$
  we have $d(e,n)=0$: termination lines of the first order transition;
  Medium dark lines e.g $\widehat{CP_m}$.:
  $\vecb{v}_1\cdot\mbox{\boldmath$\nabla$}d=0$; here the curvature
  determinant has a minimum in the direction of the largest curvature
  eigenvector $\vecb{v}_1$; in the cross-region (light gray) we have:
  $d=0\wedge\mbox{\boldmath$\nabla$}d=\mbox{\boldmath$0$}$ this is the
  locus of the multi-critical point $P_m$ where $s(e,n)$ is flat up to
  at least third order in $\Delta e$ and $\Delta n$. The two
  horizontal lines give the positions of the two cuts shown in
  figs.\ref{cut1},\ref{cut2}.
\label{det}}
\end{figure}

Figure (\ref{6d3_e}) shows $s_{micro}(e,n)$. Grid lines are in the
direction $[e-e_0(n)]/[e_{max}(n)-e_0(n)]=$ const and $n=$const,
($e_0(n)=-2n$, $e_{max}(n)=-\frac{2n^2}{q}$).  The grey levels are
determined in fig.\ref{det}: Regions above $\widehat{CP_mB}$ :
concave, $d>0$, pure phase (disordered, gas), in the triangle
$A$$P_m$$C$ concave, pure phase, ordered (solid); Below
$\widehat{AP_mB}$: convex, $d<0$, phase-separation, first order; At
the dark lines like $\widehat{AP_mB}$ we have $d(e,n)=0$: these are
termination lines of the first order transition; Medium dark lines like
$\widehat{CP_m}$: $\vecb{v}_1\cdot\mbox{\boldmath$\nabla$}d=0$, here
the curvature determinant has an extremum in the direction of the
largest curvature eigenvector $\vecb{v}_1$ (remember: in the normal
concave region of $s(e,n)$ the curvatures are both {\em negative}. A
vanishing or even small positive curvature is an abnormal large
curvature); In the cross-region (light gray without grid) we have:
$d=0\wedge\mbox{\boldmath$\nabla$}d=\mbox{\boldmath$0$}$ this is the
locus of the multi-critical point $P_m$ where $s(e,n)$ is
(numerically) flat up to at least third order in $\Delta e$ and
$\Delta n$.

The convex region of phase separation corresponds to the similar
region in the Ising lattice gas, respectively the original Ising model
as function of magnetization, c.f. the darkly shaded region in figure
(\ref{isingG}). The main new feature of the phase diagram of the diluted
($q=3$)-Potts model compared to the diluted ($q=1$)-Potts model is the
critical line from $C$ to $P_m$.

In figure (\ref{det}) the determinant of curvatures of $s(e,n)$:
\begin{equation}
d(e,n)= \left\|\begin{array}{cc}
\frac{\partial^2 s}{\partial e^2}& \frac{\partial^2 s}{\partial n\partial e}\\
\frac{\partial^2 s}{\partial e\partial n}& \frac{\partial^2 s}{\partial n^2}
\end{array}\right\|
= \left\|\begin{array}{cc}
s_{ee}&s_{en}\\
s_{ne}&s_{nn}
\end{array}\right\|=\lambda_1\lambda_2
 \label{curvdet}
\end{equation}
is shown. On the diagonal we have the ground-state of the $2$-dim
Potts lattice-gas with $e_0=-2n$, the upper-right end is the complete
random configuration (here without contour lines), with the maximum
allowed excitation $e_{rand}=-\frac{2n^2}{q}$. In the region above the
line $\widehat{CP_mB}$ we have the disordered, ``gas''.  Here the
entropy $s(e,n)$ is concave ($d>0$), both curvatures are negative (we
have always the smaller one $\lambda_2<0$).  This is also the case
inside the triangle $A$$P_m$$C$ (ordered, ``solid'' phase). In these
regions the Laplace integral eq.(\ref{grandsum}) has a single
stationary point. They correspond to pure phases.  \newpage

\begin{minipage}[t]{7cm}
\begin{figure}
  \includegraphics*[bb =0 0 290 180, angle=-0, width=7cm,
  clip=true]{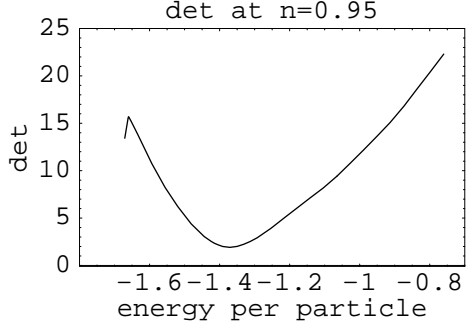}
  \\
\caption{Cut through the determinant $d(e,n)$ along 
  the line shown in figure (\ref{det}) at const.  $n=0.95$, through
the critical line $\widehat{CP_m}$ close
  to the critical point $C$ of the ordinary Potts model ($n\sim 1$)
\label{cut1}}
\end{figure}
\end{minipage}\rule{0.5cm}{0mm}\begin{minipage}[t]{7cm}
\begin{figure}
  \includegraphics*[bb =0 0 290 180, angle=-0, width=7cm,
  clip=true]{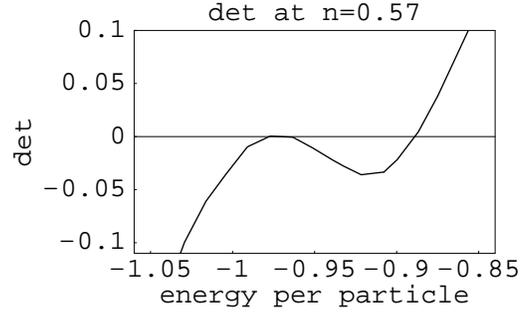}
  \\
\caption{Cut through the determinant $d(e,n)$ along 
  the line shown in figure (\ref{det}) at const.  $n=0.57$, slightly
  below the multi-critical region. There are several zero points of
  the determinant of curvatures: The left one is simultaneously a
  maximum with $\mbox{\boldmath$\nabla d=0$}$ and consequently
  critical as discussed above 
 \label{cut2}}
\end{figure}
\end{minipage}
Below $\widehat{AP_mB}$ $s(e,n)$ is convex ($d<0$) c.f. figure
(\ref{intrudbeta}), corresponding to phase-separation, first order.  At
these \{$e,n$\} the Laplace integral (\ref{grandsum}) has no
stationary point. Here we have a separation into coexisting phases,
e.g. solid and gas. Due to the intra-phase surface tension or the negative
contribution to the entropy by the additional correlations at the
phase boundaries (surface), $s(e,n)$ has a {\em convex} intruder with
positive largest curvature.  In \cite{gross150,gross157} it is shown
that the depth of the convex intruder in $s(e,n)$ gives the surface
tension.

At the dark lines like $\widehat{AP_mB}$ we have $d(e,n)=0$. These
are the termination lines of the first order transition. At these
lines one of the two phases is depleted and beyond all particles are
in the other phase (solid or gas respectively).

Along the medium dark lines like $\widehat{P_mC}$ we have
$\vecb{v}_1\cdot\mbox{\boldmath$\nabla$}d=0$, here the curvature
determinant has a minimum in the direction of the largest curvature
eigenvector $\vecb{v}_1$. The line $\widehat{P_mC}$ towards the
critical point of the ordinary ($q=3$)-Potts model at $e=-1.58$, $n=1$
correponds to a critical line of second order transition which
terminates at the multicritical ``point'' $P_m$.  It is {\em a deep
valley} in $d(e,n)$ c.f. fig.\ref{cut1} which rises slightly up
towards $C$. On the level of the present simulation we cannot decide
whether this rise is due to our still finite, though otherwise
sufficient, precision or is a general feature of finite size. (The
largest curvature $\lambda_1$ of $s(e,n)$ has a local {\em maximum}
with $\lambda_1\stackrel{<}{\sim}0$, or $d\stackrel{>}{\sim}0$).
Because of our finite interpolation width of $\Delta e\sim \pm 0.04$,
$\Delta n\sim\pm 0.02$ it might be that this valley of $d(e,n)$ gets
a little bit filled up from its sides and the minimum is rounded,
c.f. fig.\ref{cut1}.
The valley converts below the crossing point $P_m$ into {\em a flat
ridge} inside the convex intruder of the first order lattice-gas
transition see also fig.\ref{intrudbeta}.

In the cross-region (light gray in fig.\ref{det}) we have:
$d=0\wedge\mbox{\boldmath$\nabla$}d=\mbox{\boldmath$0$}$.  This is the
locus of the multi-critical point $P_m$ where $s(e,n)$ is
(numerically) flat up to at least third order in both directions
$\Delta e$ and $\Delta n$.  It is at $e_m\sim -1$, $n_m\sim 0.6$ or
$\beta_m= 1.48\pm 0.03$, $\nu_m= 2.67\pm 0.02$.  Naturally, $P_m$
spans a much broader region in \{$e,n$\} than in \{$\beta,\nu$\},
remember $s(e,n)$ is {\em flat} near $P_m$. This situation reminds
very much the well known phase diagram of a $^3$He --$^4$He mixture
in temperature vs.  mole fraction of $^3$He c.f.  fig.3. in
ref.\cite{lawrie84}.

Figure (\ref{intrudbeta}) gives a 3D-view of the entropy-surface
$s(e,n)-\beta_0 (e-e_0)-\nu_0(n-n_0)$ along a broad strip with
constant $\beta=-1.85$ (through the region of phase separation,
roughly parallel to the ground state $e_0(n)=-2n$).  The overall convex
intruder of the first order transition is well seen. In its middle the
narrow ridge where $s(e,n)$ is again concave can be imagined.  This is
the origin of the medium dark line $\widehat{P_mD}$ in figures
(\ref{6d3_e}) and (\ref{det}).
\begin{center}\begin{figure}
    \includegraphics*[bb =0 0 290 250, angle=-0,
    width=15cm,clip=true]{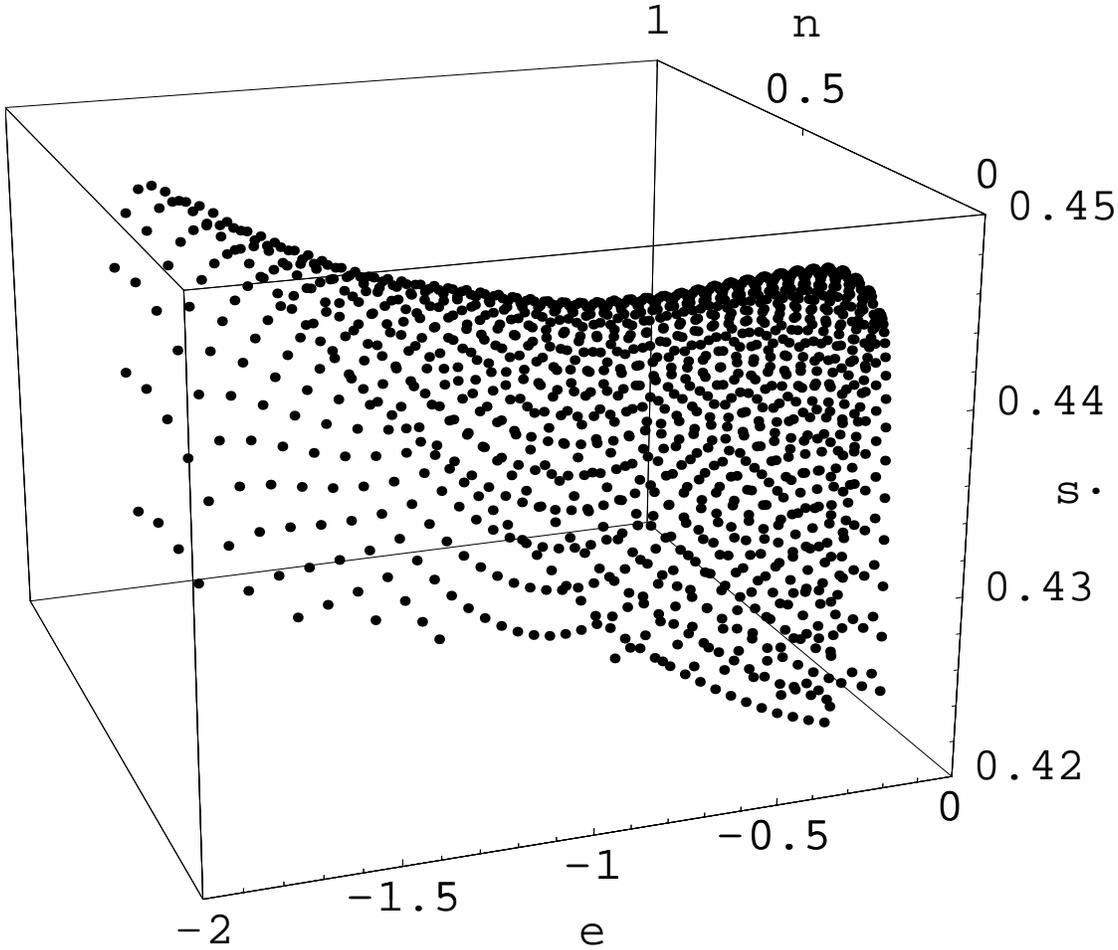}
\caption{Entropy surface $s'=s(e,n)-1.85(e-e_0)-3.61(n-n_0)$ along
  a strip with constant $\beta=1.85$ (roughly parallel to the
  ground state $e_0(n)=-2n$ through the region of phase separation).
\label{intrudbeta}}
\end{figure}\end{center} 

\subsection{On the topology of curvatures}
\begin{eqnarray}
&&\mbox{The two eigenvalues of the curvature matrix (\ref{curvdet}) are:}
\nonumber\\
\lambda_{1,2}&=&\frac{s_{ee}+s_{nn}}{2}\pm\frac{1}{2}\sqrt{(s_{ee}
+s_{nn})^2-4d}\\
&&\mbox{and the corresponding eigenvectors are :}\nonumber\\
\vecb{v}_\lambda&=&\frac{1}{\sqrt{(s_{ee}-\lambda)^2+s_{en}^2}}
\left(\begin{array}{l}
                                        -s_{en}\\
                                         s_{ee}-\lambda
                                    \end{array}\right).\\
\end{eqnarray}
At critical points the following conditions hold:
\begin{eqnarray}
d&=&-\frac{\partial(\beta\nu)}{\partial (en)}=L^2D=0\\
s_{ee}s_{nn}&=&s_{en}^2.\\
&&\mbox{Here the directions $\beta=$const. and $\mu=$ const. are {\em
parallel}}\\
&&\mbox{and we have :}\nonumber\\
\left.\frac{\partial \beta}{\partial e}\right|_{\nu}&=&\frac{d}{s_{nn}}=0\\
\left.\frac{\partial \nu}{\partial n}\right|_{\beta}&=&\frac{-d}{s_{ee}}=0.\\
\lambda_1&=&0\\
\lambda_2&=&s_{ee}+s_{nn}\\
\vecb{v}_{\lambda=0}&=&\frac{1}{\sqrt{s_{ee}^2+s_{en}^2}}\left(\begin{array}{l}
                                        -s_{en}\\
                                         s_{ee}
                                    \end{array}\right)\\
\vecb{v}_{\lambda\le0}&=&\frac{1}{\sqrt{s_{nn}^2+s_{en}^2}}\left(
\begin{array}{l}
s_{en}\\ s_{nn} \end{array}\right).\end{eqnarray}

The vanishing of $d$ is not sufficient for criticality.  Physically,
it means that the surface entropy (tension) and with it the interface
separating coexistent phases disappears. This, however, can also
signalize a depletion of one of the two phases.  At a critical
end-point, however, the interface disappears at a non vanishing
number of atoms in each of the two phases.  I.e.  in an infinitesimal
neighborhood of a critical point, $d$ must remain zero.  In a
topologically formulation a critical end-point of first
order transition is at:
\begin{eqnarray}
                                   d&=&0\\
                                   &\mbox{\em and}&\nonumber\\
                                   \vecb{v}_1\cdot\mbox{\boldmath$\nabla$}
                                   d&=&0 .
\end{eqnarray} 

This is a generalization of the well known condition for a continuous
transition in one dimension: the simultaneous vanishing of
$\beta^\prime(e)=0$ and of the curvature of $\beta(e)$,
$\beta^{\prime\prime}(e)=0$.

\begin{center}\begin{figure}
    \includegraphics*[bb =0 0 290 180, angle=-0,
    width=13cm,clip=true]{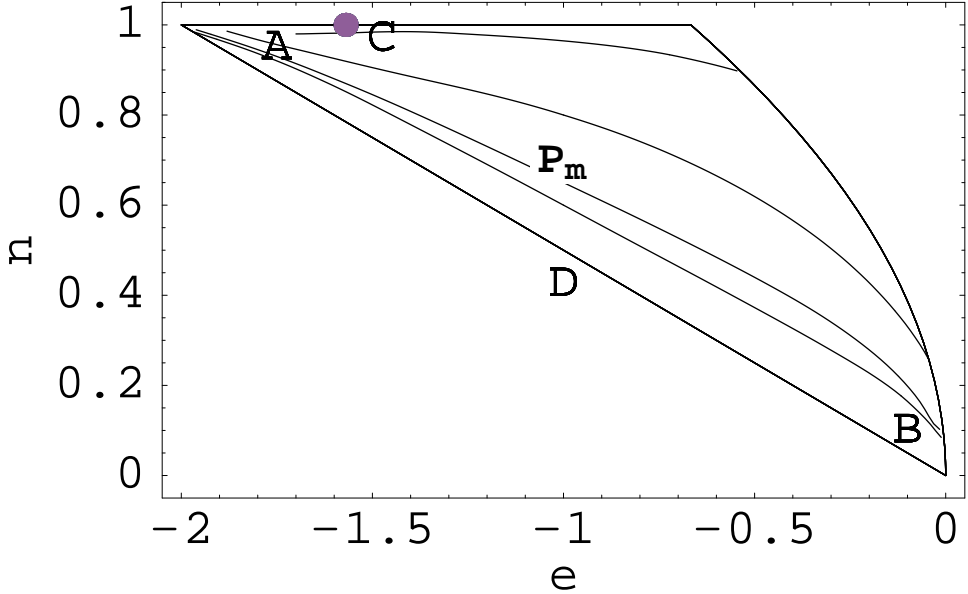}
\caption{Direction of the largest principal curvature $\mbox{\boldmath$v_1$}$ .
\label{maincurvature}}
\end{figure}\end{center} 
Figure (\ref{maincurvature}) shows a map of some trajectories which
follow the eigen-vector $\mbox{\boldmath$v_1$}$ with the largest
curvature eigen-value $\lambda_1$. In the region of the convex
intruder ($\lambda_1>0$) i.e. the region of phase-separation
$\mbox{\boldmath$v_1$}$ is $\sim$ parallel to the ground state
$e=-2n$. Also the lines of $\beta=$const. and $\mu=$const. follow
approximately this direction. Their Jacobian
$\partial(\beta\nu)/\partial(en)=d(e,n)$ is negative but small.  This
reminds of the situation in the thermodynamic limit where this region
of phase coexistence is flat, both intensive variables are constant
and the Jacobian $d\to-0$.  One can also see in fig.\ref{maincurvature}
how the direction of the largest curvature $\mbox{\boldmath$v_1$}$
turns into the $e$-direction when one approaches the critical point
$C$ of the ordinary ($q=3$)-Potts model at $n=1$.

\begin{figure}
  \includegraphics*[bb =0 0 375 270, angle=-0, width=11cm,
  clip=true]{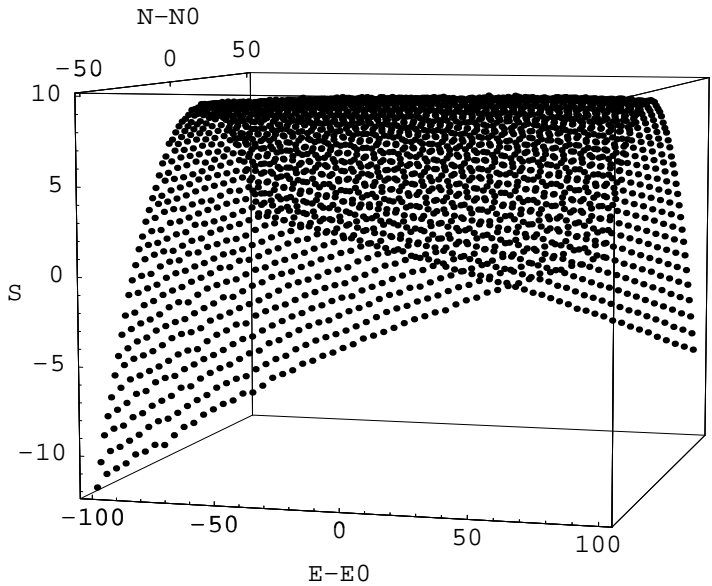}
  \\
\caption{Histogram of $s(e,n)-(e-e_0)\beta(e_0,n_0)-(n-n_0)
  \nu(e_0,n_0)$ at $e_0=-1.48,n_0=0.94$ in the rectangle $e_0\pm
  0.04$, $n_0\pm 0.02$ viewed in the direction of the valley
  $\widehat{P_mC}$ of $d(e,n)$, the second-order critical line towards
  the critical point $C$ of the ordinary $q=3$-Potts model at
  $e=-1.58$. The approximate vanishing of the largest curvature in the
  direction $\vecb{v}_1$ (pointing from left to right) at this point
  $e_0,n_0$ is clearly seen.
  \label{histogr}}
\end{figure}

Inspection of fig.\ref{det} shows that along the line $\widehat{CP_m}$
of second order transition $\vecb{v}_1\cdot\mbox{\boldmath$\nabla$}
d=0$, i.e.  $d(e,n)$ has a deep and sharp valley. This line is the
locus of a minimum of $d$ in the direction of the largest eigenvalue
$\vecb{v}_1$ of the curvature. In the direction of $\widehat{CP_m}$
the determinant $d(e,n)$ is slightly growing towards the ordinary
Potts critical point $C$. Figure (\ref{histogr}) shows an original
narrow histogram of $s(e,n)$, ($\Delta e=0.08$ and $\Delta n=0.04$) on
the line $\widehat{CP_m}$ near to the point $C$ which shows that the
curvature of $s(e,n)$ is rather small in the direction of the larger
curvature eigenvector $\vecb{v}_1$.

At $n=1$ we know that for an infinite system the ordinary ($n=1$)
three state Potts model has a second order transition at $e=-1.58$
where the curvature of $s(e)$ vanishes, $s_{ee}=0$. I.e. the component
$\vecb{v}_1\cdot\mbox{\boldmath$\nabla$} d$ of
$\mbox{\boldmath$\nabla$} d$ indicates nicely the locus of the second
order ``temperature driven'' transition of the ordinary Potts model.
\subsection{The information lost in the grand-canonical ensemble
\label{lostinfo}}
\begin{figure}
\includegraphics*[bb = 0 0 290 280, angle=-0, width=13cm,
  clip=true]{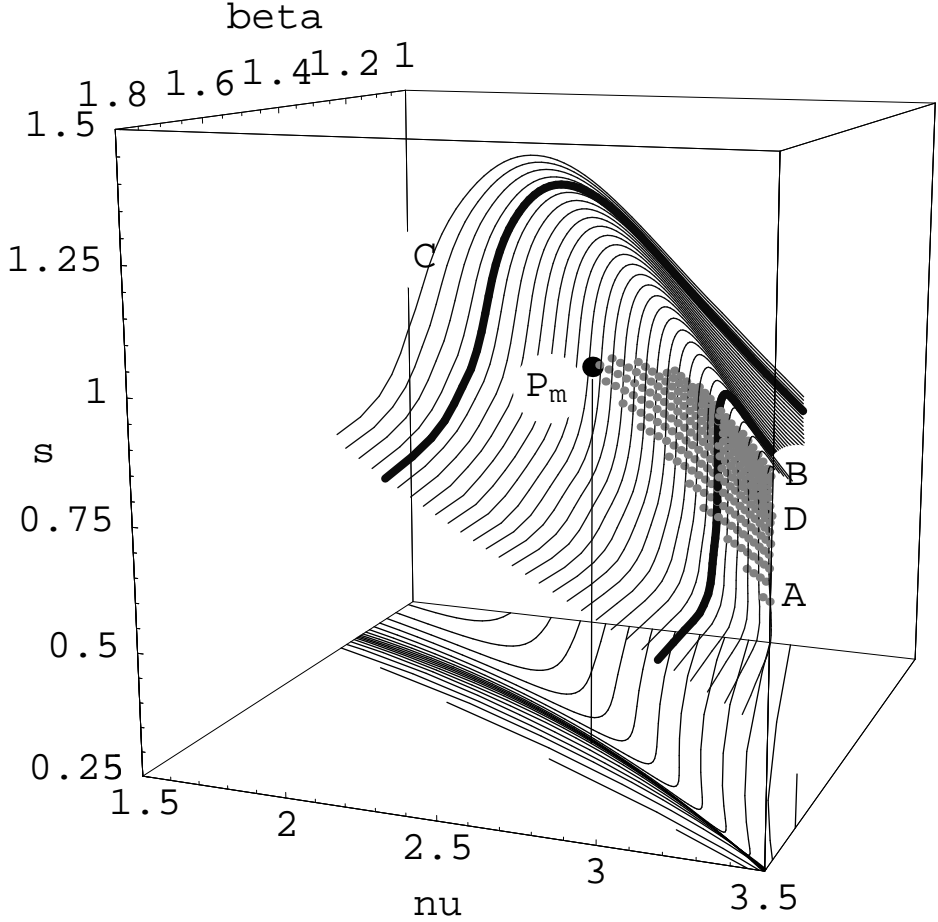}
\caption{Plot of the entropy $s_{micro}(\beta,\nu)$ as function of
  the ``intensive'' variables ($\nu=-\beta\mu$) in the figure labeled
  as ``nu'' and $\beta$ is called ``beta''.  The lines which build the
  surface are lines for $\beta=$const. The two bold ones indicate the
  cuts shown in figs.\protect\ref{sbetaless} and
  \protect\ref{sbetahigh}.  The positions of the points
  $A$,$D$,$B$,$C$ defined in figure (\ref{det}) are only roughly
  indicated. The convex intruder where $s_{micro}(\beta,\nu)$ becomes
  multi-valued as function of $\nu>\nu_{P_m}$ and $\beta>\beta_{P_m}$
  is indicated by shadowing. This corresponds to the first order
  transition. At the bottom the projection of the entropy surface onto
  the \{$\beta, \nu$\} plane is shown as contour plot. The convex part
  (region of phase-separation) is hidden behind the dark ``critical''
  line.
\label{Sintens}}
\end{figure}
Figure (\ref{Sintens}) explains what happens if one plots the entropy
$s$ vs. the ``intensive'' quantities $\beta=\partial S/\partial E$ and
$\nu =\partial S/\partial N$ as one would do for the grand-canonical
ensemble: As there are several points $E_i,N_k$ with identical
$\beta,\nu$, $s_{micro}(\beta,\nu)$ is a multivalued function of
$\beta,\nu$. Here the entropy surface $s_{micro}(e,n)$ is folded onto
itself. In the projection in fig.\ref{Sintens}, these points show up
as a black critical line (dense region). Here this black line continues
over the multi-critical point $P_m$ towards $C$ indicating the
direction to the critical point of the ordinary $q=3$ Potts model at
$n=1$ (zero vacancies). Between $P_m$ and $C$ the slopes
\begin{eqnarray}
\left.\frac{\partial s}{\partial \beta}\right|_\nu&=&
\frac{1}{d}[\beta s_{nn}-\nu s_{ne}]\\
\mbox{or}&&\\
-\left.\frac{\partial s}{\partial \nu}\right|_\beta&=&
\frac{1}{d}[\beta s_{en}-\nu s_{ee}]
\end{eqnarray}
are negative large but finite.

The information given by the projection would be all information which
can be obtained from the conventional grand-canonical entropy
$s(T,\mu,V)$, if we would have calculated it from the Laplace
transform, eq.(\ref{grandsum}). The shaded region will be lost.

The upper part of figure (\ref{Sintens}) shows $s_{micro}(\beta,\nu)$
in a three dimensional plot. The lines building the entropy surface are
lines of equal $\beta$. The images of the points $A,D,B,C$ defined in
fig.\ref{det} are roughly indicated.  The back folded branches, the
convex intruder of $s(e,n)$ between the lines $\widehat{AP_mB}$ and
$\widehat{ADB}$, the region of phase separation, can here be seen from
the side (shadowed). It is jumped over in eq. (\ref{grandsum}) and
gets consequently lost in $Z(T,\mu)$. This demonstrates the far more
detailed insight into phase transitions and critical phenomena
obtainable by micro-canonical thermo-statistics which is not
accessible by the canonical treatment, c.f. the similar arguments of
Gibbs \cite{gibbs06}.

\begin{figure}
  \includegraphics*[bb = 0 0 290 180, angle=-0, width=10cm,
  clip=true]{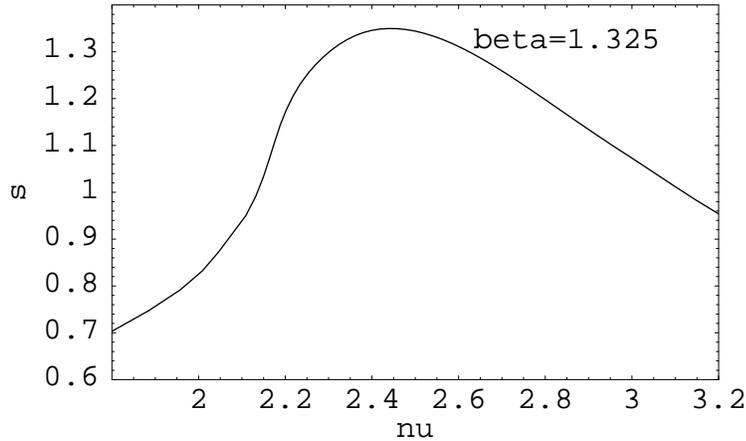}
\caption{Plot of the entropy $s(\beta=1.325,\nu)$\label{sbetaless}}
\end{figure}

\begin{figure}
  \includegraphics*[bb = 0 0 290 180, angle=-0, width=10cm,
  clip=true]{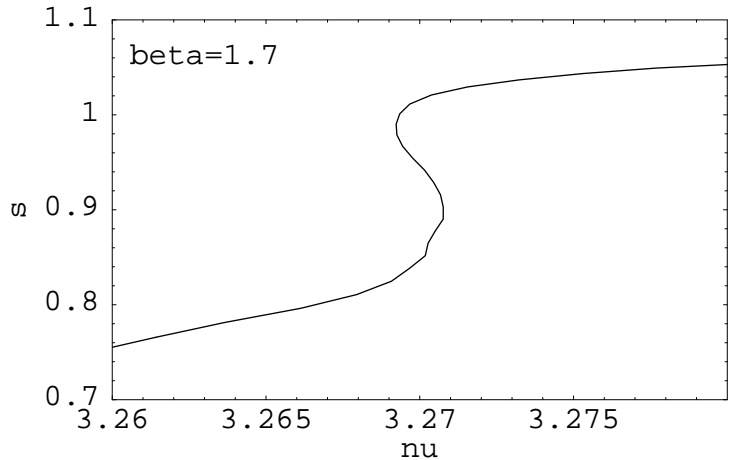}
\caption{Plot of the entropy $s(\beta=1.7,\nu)$\label{sbetahigh}}
\end{figure}

In the next two figures the cross-section through $s(\beta,\nu)$
at constant $\beta$ is shown in figure (\ref{sbetaless}) below the
multi-critical point $\beta_m=1.48$ and in figure (\ref{sbetahigh})
above it.  The latter clearly shows the back-bending of
$s(\beta,\nu)$.
\section{Convex entropy --- Violation of the Second Law ?}
At this point it is worth-wile to spend some words on a popular
misunderstanding connected with the eventual convexity of the entropy
as function of ``extensive'' quantities like the energy: The convex
parts of $S(E,N)$ violate van Hove's concavity condition
\cite{vanhove49,hill55}.

One may believe that this is also in contradiction to the second law
of thermodynamics: At a convex region of $S(E,N)$ a split of the
system into two pieces with entropies $S_1(E_1,N_1)$ and
$S_2(E_2,N_2)$ would have
$S_1(E_1,N_1)+S_2(E_2,N_2)>S(E_1+E_2,N_1+N_2)$. So the system seems to
{\em gain} entropy by splitting.

This, however, is an error. The Boltzmann entropy as defined in the
section II is already the logarithm of the sum over {\em all} possible
configurations of the system at the given energy. The split ones are a
subset of these. Their partial phase space $W_{split}$ is of course
$\le$ the total $W$. The entropy $S_{split}=ln(W_{split})$ is $\le$
the total entropy. Evidently, the split system looses some surface
entropy $S_{surf}$ at the separation boundary due to additional
correlations imposed on the particles at the boundary, see the
discussion in section \ref{surf}.  The entropy after split is
consequently:
\begin{eqnarray}
S_{split}&=&S_1(E_1,N_1)+S_2(E_2,N_2)-S_{surf}\nonumber\\
&\le&S(E_1+E_2,N_1+N_2),
\end{eqnarray}
It is a typical finite size effect. $S_{surf}/V$ vanishes in the limit
$V\to\infty$ for interactions with finite range. The entropy is
non-extensive for finite systems but becomes extensive in the limit,
and van Hove's theorem \cite{vanhove49} is fulfilled. This is of
course only under the condition that $\lim_{V\to\infty}S_{surf}/V =0$.

In general this is of course a trivial conclusion: An additional
constraint like an artificial cut of the system can only reduce phase
space and entropy. The Second Law is automatically satisfied in the
Boltzmann formalism whether $S$ is concave or not, whether $S$ is
``extensive'' or not.

A positive (wrong) curvature introduces problems to the geometrical
interpretation of thermodynamics as formulated by Weinhold
\cite{weinhold75,weinhold78} which relies on the non-convexity of
$S(E,N)$.  Weinhold introduces a metric like
\begin{eqnarray}
g_{ik}&=&-\frac{\partial^2S}{\partial X^i\partial X^k}\\
&&\mbox{where we identify :}\nonumber\\
X^1&=&E\nonumber\\
X^2&=&N .\nonumber\\
&&\mbox{The thermodynamic distance is defined as :}\nonumber\\
\Delta_{a,b}&=& \sqrt{[X^i(a)-X^i(b)]g_{ik}[X^k(a)-X^k(b)]}.
\end{eqnarray} 
Evidently, a negative metric $g_{ik}$ is here not allowed. Of course
Weinhold's theory does not apply to finite systems with phase
transitions.
\section{Conclusion} 
Micro-canonical thermo-statistics describes how the entropy $s(e,n)$
as defined entirely in mechanical terms by Boltzmann depends on the
conserved ``extensive'' variables: energy $e$, particle number $n$,
angular momentum $L$ etc. It is well defined for finite systems
without invoking the thermodynamic limit.  Thus in contrast to the
conventional theory, we can study phase transitions also in ``small''
systems or other non-extensive systems. In this simulation we could 
classify phase transitions in a ``small'' system by the topological
properties of the determinant of curvatures $d(e,n)$,
eq.(\ref{curvdet}) of the micro-canonical entropy-surface $s(e,n)$:
\begin{itemize}
\item A single stable phase by $d(e,n)>0$.
\item A transition of first order with phase separation by $d(e,n)<0$.
  The depth of the intruder is a measure of the intra-phase surface
  tension \cite{binder82,gross150}. This region is bounded by a line
  with $d(e,n)=0$. On this line $P_m$ is a critical end-point where
  additionally $\vecb{v}_1\cdot\mbox{\boldmath$\nabla$} d=0$ in the
  direction of the eigenvector of $d(e,n)$ with the largest eigenvalue
  $\lambda_1$.
\item There, the transition is continuous (``second order'') with vanishing
  surface tension, and no convex intruder in $s(e,n)$. Here two
  neighboring phases become indistinguishable, because there are no
  interfaces. However, we found a further {\em line}
  ($\widehat{P_mC}$, critical)  with
  $\vecb{v}_1\cdot\mbox{\boldmath$\nabla$} d=0$ which does not border a
  region of negative $d(e,n)$.  Presumably $d(e,n)$ should be $0$ also.
  This needs further tests in other systems. It may also be that these
  lines signalize transitions of first order in other, but hidden
  conserved degrees of freedom.
\item Finally a multi-critical point $P_m$ where more than two phases become
  indistinguishable by the branching of several lines with $d=0$ or
  with $\vecb{v}_1\cdot\mbox{\boldmath$\nabla$} d=0$ to give a {\em
    flat} region with additionally $\mbox{\boldmath$\nabla$}
  d=\mbox{\boldmath$0$}$.
\end{itemize}

Our classification of phase transitions by the topological structure
of the micro-canonical Boltzmann entropy $s(e,n)$ is close to the
natural experimental way to identify phase transitions of first order
by the inhomogeneities of phase separation boundaries. This is
possible because the micro-canonical ensemble does not suppress
inhomogeneities in contrast to the grand-canonical one, as was
emphasized already by Gibbs \cite{gibbs06}.  Intra-phase boundaries are
reflected in ``small'' systems by the convex intruder in the entropy
surface. With this extension of the definition of phase transitions
to ``small'' systems there are remarkable similarities with the
transitions of the bulk.  Moreover, this definition agrees with the
conventional definition in the thermodynamic limit (of course, in the
thermodynamic limit the largest curvature $\lambda_1$ approaches $0$
from above at phase transitions of first order). The region of phase
separation remains inaccessible in the conventional grand-canonical
ensemble.

We believe, however, that the various kind of transitions discussed
here have their immediate meaning in ``small'' and non-extensive
systems independently whether they are the same in the thermodynamic
limit (if this then exist) or not. For systems like the Potts model
that have a thermodynamic limit it might well be possible that the
character of the transition changes towards larger system size.

The great conceptual clarity of micro-canonical thermo-statistics
compared to the grand-canonical one is clearly demonstrated.  Not only
that, we showed that the micro-canonical statistics gives more
information about the thermodynamic behaviour and more insight into the
mechanism of phase transitions than the canonical ensemble: About half
of the whole \{E,N\} space, the intruder of $S(E,N)$ or the region
between the ground state and the line $\widehat{AP_mB}$ in figure
(\ref{det}), gets lost in conventional grand-canonical thermodynamics.
Without any doubts this contains the most sophisticated physics of
this system. We emphasized this point already in \cite{gross170}
there, however, with still limited precision. Due to our refined
simulation method this could be demonstrated here with uniformly good
precision in the whole \{$E,N$\} plane. Finally, we should mention
that micro-canonical thermo-statistics allowed us to compute phase
transitions and especially the surface tension in realistic systems
like small metal clusters \cite{gross157}. Our finding clearly
disproves the pessimistic judgement by Schr\"odinger
\cite{schroedinger44} who thought that Boltzmann's entropy is only
usefull for gases.  A recent application of micro-canonical
thermo-statistics to thermodynamically unstable, collapsing systems
under high angular momentum is found at \cite{laliena98}.
\\~\\
\noindent Acknowledgment: D.H.E.G. thanks M.E.Fisher for the
suggestion to study the Potts-3 model and to test how the
multicritical point is described micro-canonically. We thank H.
Jaqaman for critical reading. We are gratefully to the DFG for
financial support.

\end{document}